\documentclass[final,3p,times,twocolumn]{elsarticle}

\usepackage{amssymb}
\usepackage{amsmath}
\usepackage{amsthm}
\usepackage{color}

\usepackage{lineno}
\newcommand{\tabincell}[2]{\begin{tabular}{@{}#1@{}}#2\end{tabular}}
\journal{Nucl. Instr. Meth. A}

\begin{document}

\begin{frontmatter}

\title{The Flash ADC system and PMT waveform reconstruction for the Daya Bay Experiment}

\author[ihep,ucas] {Yongbo Huang}
\author[ihep,lab]{Jinfan Chang}
\author[ihep] {Yaping Cheng}
\author[ihep,ucas]{Zhang Chen}
\author[ihep,lab]{Jun Hu}
\author[ihep,lab]{Xiaolu Ji}
\author[ihep,lab]{Fei Li}
\author[ihep,ucas,lab]{Jin Li}
\author[ihep,lab]{Qiuju, Li}
\author[bnl] {Xin Qian}
\author[ihep] {Jetter Soeren}
\author[sysu]{Wei Wang}
\author[ihep,lab]{Zheng Wang}
\author[sysu]{Yu Xu}
\author[ihep]{Zeyuan Yu \corref{cor1}}
\ead{yuzy@ihep.ac.cn} \cortext[cor1]{Corresponding author. Tel:+86-10-8823-6256.}

\address[ihep]{Institute of High Energy Physics, Beijing 100049, China}
\address[ucas] {University of Chinese Academy of Sciences, Beijing 100049, China}
\address[lab]{State Key Laboratory of Particle Detection and Electronics, Beijing 100049, China}
\address[sysu] {Sun Yet-Sen University, Guangzhou 510275, China}
\address[bnl] {Physics Department, Brookhaven National Laboratory, Upton, NY, USA}
\begin{abstract}
To better understand the energy response of the Antineutrino Detector (AD), the Daya Bay Reactor Neutrino Experiment installed a full Flash ADC readout system on one AD that allowed for simultaneous data taking with the current readout system. This paper presents the design, data acquisition, and simulation of the Flash ADC system, and focuses on the PMT waveform reconstruction algorithms. For liquid scintillator calorimetry, the most critical requirement to waveform reconstruction is linearity.  Several common reconstruction methods were tested but the linearity performance was not satisfactory. A new method based on the deconvolution technique was developed with 1\% residual non-linearity, which fulfills the requirement. The performance was validated with both data and Monte Carlo (MC) simulations, and 1\% consistency between them has been achieved.

\end{abstract}


\end{frontmatter}


\section{Introduction}
\label{Introduction}

The Daya Bay Reactor Neutrino Experiment provided the world's most precise measurement of sin$^22\theta_{13}$ and the effective squared mass splitting $|\Delta$m$^2_{ee}|$\ \cite{DYB-Oscillation-1230days}. A precise measurement of the reactor anti-neutrino spectra was also reported~\cite{DYB-Reactor}. Both analyses require a good understanding of the Antineutrino Detector's (AD) energy response, that is the relationship between the energy of the $\bar{\nu}_{e}$ and the reconstructed energy of the positron, the final-state lepton of the Inverse Beta Decay (IBD) reaction (Eq.~\ref{eq:Ibd}).

\begin{equation}
 \label{eq:Ibd}
 \bar{\nu}_{e} + p = e^{+} + n
\end{equation}

Liquid scintillator (LS) has been used in calorimeters for several decades. It is also utilized by Daya Bay as the $\bar{\nu}_{e}$ target and detector. The positron from an IBD reaction deposits its kinetic energy then annihilates with an electron creating two $\gamma$-rays in the LS.  The deposited energy is converted to scintillation light.  It has been established that scintillation light production is not linear with the deposited energy, due to ionization quenching and contributions from Cherenkov light\cite{Birk}, referred to as LS non-linearity.

The scintillation light is detected by photomultiplier tubes (PMTs).  The PMT output analog signal is processed by custom front-end electronics designed by the Daya Bay collaboration, giving the integrated charge and a time stamp. Daya Bay utilizes a CR-(RC)$^4$ shaping circuit to integrate the PMT signal that possesses a complex interplay with the scintillation timing profile, which consists of three exponential decayed components: a fast one with a time-constant of less than 10 ns, a medium one with a time-constant of about 30 ns, and a slow one with a time-constant larger than 150 ns. This interplay creates a non-linear relationship between the PMT's received photons and the integrated charge, referred to as electronic non-linearity, which is about 10\% from 1 p.e. to 10 p.e. for Daya Bay~\cite{DYB-Oscillation-1230days}.

Coupling between the particle-dependent LS non-linearity and energy-dependent electronic non-linearity introduces difficulties to the precise understanding of the detector energy response. To de-couple them, a Flash Analog Digital Convertor (FADC) readout system was installed for AD1 in the Daya Bay Experiment Hall.  Simultaneous data taking with the current readout system began in Feb.~2016.

The PMT's waveform recorded by the FADC in principle allows for the precise reconstruction of PMT charge and a measurement of the electronics non-linearity. However, due to the coupling of Daya Bay PMTs' overshoot and the LS scintillation timing, the precise charge extraction is not trivial. The scientific goal requires the non-linearity of PMT charge reconstruction to be no more than 1\%, which controls the size of electronics non-linearity to less than 0.3\%.

This paper presents the design, data acquisition (DAQ), and simulation of the FADC system. Several waveform reconstruction methods were developed and tested, but the 1\% linearity requirement was not satisfied initially.  Then a fast, robust and precise charge reconstruction algorithm was developed based on the deconvolution technique with 1\% non-linearity as estimated from MC.  To validate the results, data and MC were compared achieving agreement at 1\% level.

\section{Flash ADC system}
\label{FADC}
\subsection{System design}

In Daya Bay, each AD contains 192 Hamamatsu R5912 8'' PMTs, operating under positive high voltages (HV). The HV and PMT output signal share the same cable, and to decouple them, a custom-built splitter filters the high voltage.  The analog signal is passed to the front-end electronics (FEE).  Details of the current readout system are reported elsewhere\ \cite{DYB-Detector}.

\begin{figure}[!htb]
\begin{centering}
\includegraphics[width=.5\textwidth]{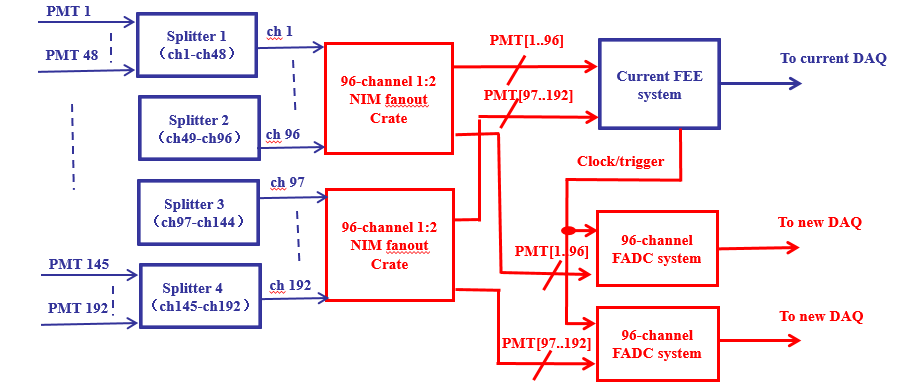}
\caption{\label{fig:FADCSystem} Signal distribution scheme to the Daya Bay FADC system. PMT signals were sent into a FIFO with two outputs. One output was fed into the old electronics named FEE, and the other one with 10 times amplification was fed into the new FADC system. }
\end{centering}
\end{figure}

The signal distribution scheme to the newly installed FADC system is shown in Fig.~\ref{fig:FADCSystem}.

To enable FEE and FADC simultaneously data taking, a Fan-In-Fan-Out (FIFO) system with 192 channels was built. The PMT signals from the splitter are fed into the FIFO with two outputs. One output with 1x amplification is sent to the FEE, and the other one with 10x amplification is sent to the FADC. The chip used in the first output (for the FEE) is ADA4817, with a -3 dB bandwidth 1.05 GHz.  The second output uses the AD8000, chip with a -3 dB bandwidth 1.5 GHz. Both amplifiers are fast enough to prevent distortions, since the signals concentrated in the frequency range less than 300 MHz.

\begin{figure}[!htb]
\begin{centering}
\includegraphics[width=.3\textwidth]{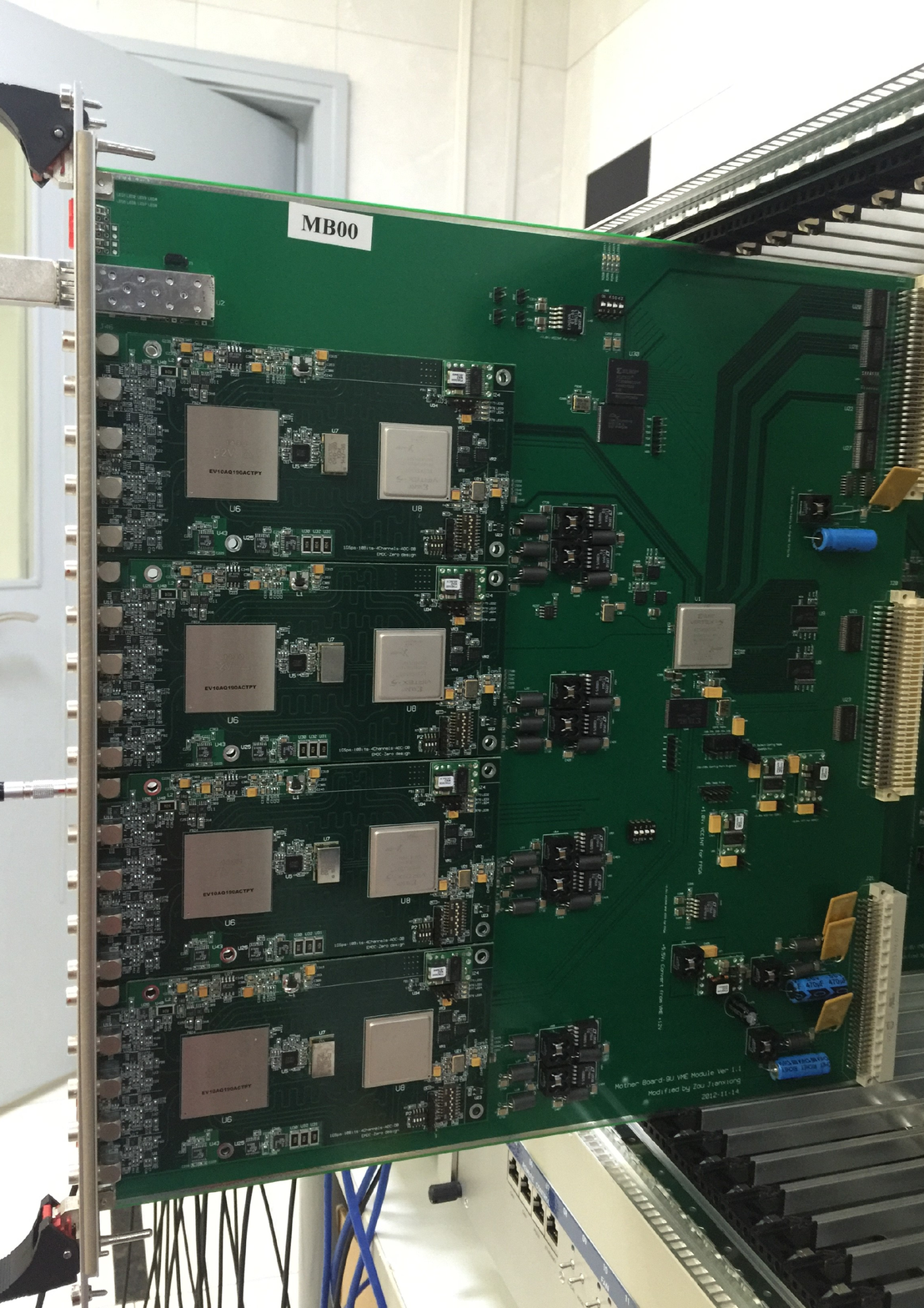}
\caption{\label{fig:FADCBoard} Photograph of one FADC board consisting of 16 channels. The mother board is a 9U-VME board. It collects the raw data from 4 daughter boards.}
\end{centering}
\end{figure}

The custom-built FADC system consists of twelve boards with 192 channels in total. The ADC chip operating with a sampling rate of 1 GHz, an effective number of bits (ENOB) of 7.8, and a maximum range of 500 mV. The main function of the high speed ADC module is signal conditioning, high speed AD conversion, data capture and transmitting the interesting data to the DAQ according to the trigger signal. The module is based on the daughter-mother board structure and both boards are 8 layers pcb. The structure of this module is shown in Fig.~\ref{fig:FADCBoard}. The daughter board is based on a 4-channel, 1 GSps, 10-bit ADC chip and the model is EV10AQ190.

To provide high performance 1 GHz clock to ADC, a HMC830LP6GE jitter cleaner is used. The analog signal from the detector is converted into a differential signal through a high speed amplifier and then transferred to the ADC chip. The FPGA captures the ADC’s output digital data. After receiving a trigger signal, the FPGA packages the interesting data and transmits it to DAQ server.

The mother board is a 9U-VME board. It collects the raw data from 4 daughter boards and transmits them to the data acquisition system via gigabit Ethernet with TCP/IP protocol. It also receives the trigger signal and fans it out to all the ADC daughter boards. A remote update circuit based on CPLD makes updating the firmware of the system convenient. As the key component in this system, FPGA is Xilinx XC5VSX50T which contains enough logic, ram and interface resource for the data transmitting, especially DSP resource for the complex signal processing algorithm in the future.

The readout window length is configurable from 100 ns to 100 $\mu$s, and was chosen to be 1008 ns during data taking. The window length is chosen to ensure the overshoot is well recovered while minimizing the data volume.

The trigger of FADC and FEE comes from the same trigger board. Trigger signal from the trigger board goes to a Fan-In-Fan-Out (FIFO) board which is installed on the crate of FEE. The FIFO has more than ten outputs which are sent to each FEE board and FADC board. And the cables connecting the FIFO and FEE/FADC have the same length. So the FADC system shared the same trigger and global time with the FEE. Both FEE and FADC events contain global trigger time stamps which could be used to align the events recorded by two systems.

\begin{figure}[!htb]
\begin{centering}
\includegraphics[width=.45\textwidth]{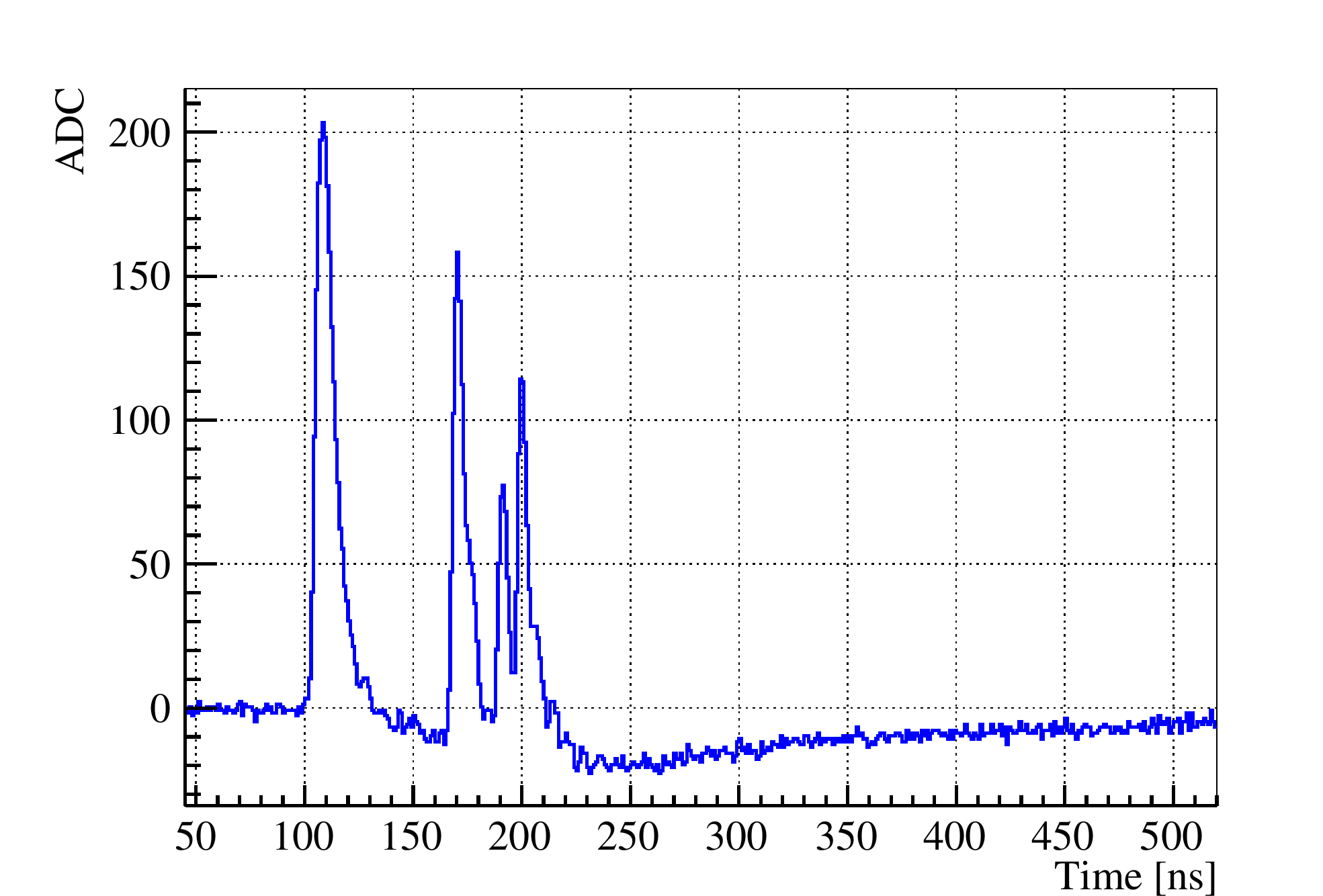}
\caption{\label{fig:WaveformExample} One PMT waveform with pile-up hits from data. Amplitude of a single p.e. was about 80 ADC counts (40 mV). The overshoot came from the PMT base and the HV-signal splitter. }
\end{centering}
\end{figure}

\begin{figure}[!htb]
\begin{centering}
\includegraphics[width=.45\textwidth]{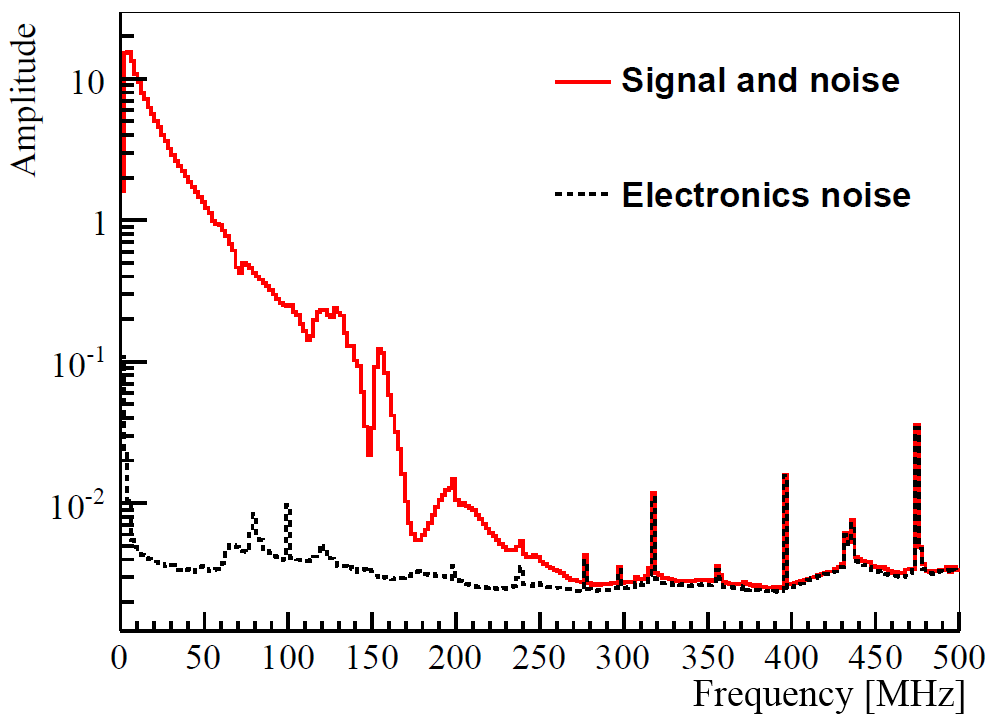}
\caption{\label{fig:FrequencyAna} Frequency response of the FADC data. The red solid line is the sum of PMT signals and noise, and the black dash line is pure electronics noise. }
\end{centering}
\end{figure}

An example waveform taken by the FADC system is shown in Fig.~\ref{fig:WaveformExample}. The amplitude of a single p.e. is about 40 mV (about 80 ADC counts) after 10x amplification. Frequency analysis shows that signals concentrated to less than 300 MHz, which is due to the PMT itself, and the electronics noise was almost white, as shown in Fig.~\ref{fig:FrequencyAna}. It can be concluded that the FADC design fulfilled the Sampling Theorems\ \cite{Sampling} and Quantisation Theorems\ \cite{Quantisation}.

Since the Daya Bay experiment utilizes the near and far relative measurement to measure the neutrino mixing angle $\theta_{13}$, it was critical to prevent FEE response differences before and after the FADC installation.  Many checks were done to demonstrate that the FEE was working as before, including PMT gains, dark rates, measured charge and time spectra, etc. Nothing abnormal was found, indicating that the relative $\bar{\nu}_{e}$ detection efficiencies between AD1 and the other ADs was not affected.

\subsection{Data acquisition}

Although benefiting from a high sampling rate, huge data volumes are a significant issue for experiments employing FADCs. Data of the FADC system at Daya Bay were sent to the new DAQ via Ethernet, the raw trigger rate of the FADC system was about 240 Hz, generating 80 MB/s of raw data, which was much larger than the 20 MB/s network bandwidth between the experiment site and the offline computing center. An effective data reduction strategy was therefore developed.

The DAQ scheme and its implementation was reported elsewhere\ \cite{DYB-DAQ}. The DAQ program was further developed for the data reduction which consisted of three steps~\cite{DAQReduction}.

1) Time correlation cut. The DAQ recorded only event pairs within a time interval less than 0.5 ms. This eliminates most of the single events from natural radioactivity, keeping the events of interest, for example, Inverse Beta Decay and Bi-Po cascade decays. After this step, the trigger rate and data size are decreased to 30 Hz and 10 MB/s, respectively.

2) Energy cut. A prompt energy reconstruction was done in the DAQ program. Events with an energy less than 0.6 MeV or larger than about 15 MeV are removed. The trigger rate and data size were further reduced to 13 Hz and 5 MB/s.

3) Empty channel cut. In each event, if a channel did not pass a 0.3 p.e. threshold, it was removed, resulting in a 2 MB/s data size.

The time correlation cut is done on the FADC board, DAQ program doesn't read events which don't pass this cut. The other cuts are done in the memory of DAQ server, before writing data to hard disk. With this effective data reduction algorithm, the data size fulfilled the requirements for the data transferring from Daya Bay to the computing center.

\section{FADC simulation}
\label{Simulation}

To model the FADC responses and validate the PMT charge reconstruction algorithms, a precise single channel electronics simulation was developed, with the following steps.

1) Sample the PMT hit number and time. The hit number distribution was user defined, i.e. physical or uniformly distributed. The time of each hit was sampled from the measured LS scintillation timing profile.

2) Convolve the hits with analog PMT single p.e. waveform, which was modeled by a data-driven function, to be discussed in Sec.~\ref{SPEModel}. The amplitude of single p.e. followed a Gaussian distribution with 30\% resolution according to the real PMT response.

3) Add analog electronics noise and the baseline offset. White noise with 0.7 mV sigma and measured noise from Daya Bay FADC data were tested, and no differences were found.

4) Digitization. The analog waveform was digitized with the FADC configurations, following the round off principle.

\subsection{Single p.e. waveform modeling}
\label{SPEModel}

\begin{figure}[!htb]
\begin{centering}
\includegraphics[width=.48\textwidth]{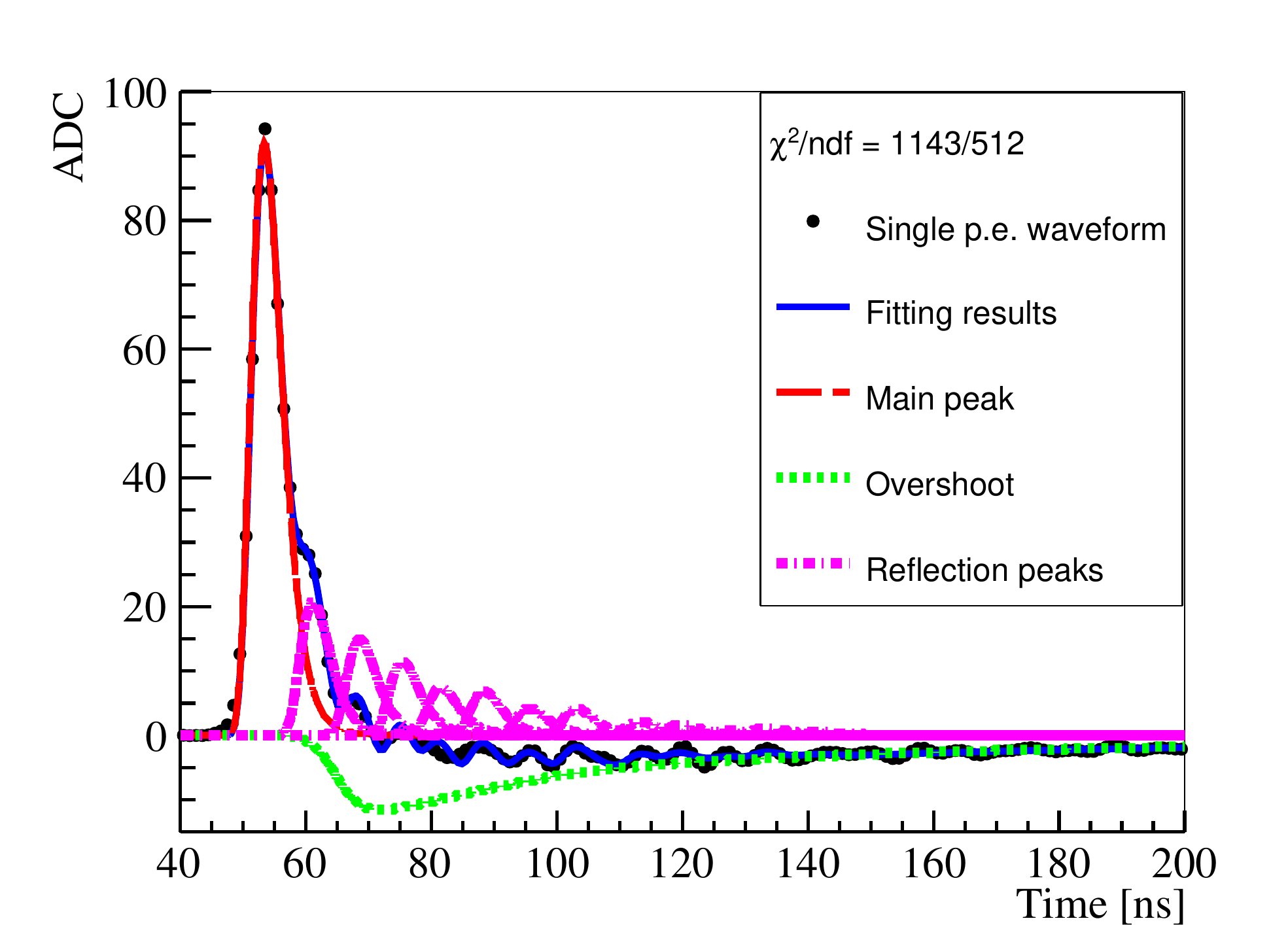}
\caption{\label{fig:SPEWaveform} One PMT's single p.e. waveform from data (black points), and the fitting results (blue solid line). The waveform consists of the main peak (red dash line), overshoot (greed dot line), and reflections (magenta dash-dot line).}
\end{centering}
\end{figure}

The waveform modeling of Hamamatsu R5912 PMTs was reported elsewhere\ \cite{PreviousModel}, in which the PMT waveform was measured in the laboratory. To better describe the single p.e. shape, we improved the model based on the Daya Bay FADC data.  One PMT's averaged single p.e. waveform measured by FADC is shown in Fig.~\ref{fig:SPEWaveform}, and it consists of three components: the main peak, the overshoot, and twelve reflection peaks.

The shape of the main peak was modeled with the log-norm function as in Eq.~\ref{eq:log-norm}. The origin of reflections was unknown and the time interval between them was about 7 ns. Empirically they were modeled with the same shape as the main peak, but with different amplitude $U^0_{peak}$ and starting time $t_0$. The overshoot was modeled with three components, a Gaussian one, a fast exponential decay and a slow exponential decay, as shown in Eq.~\ref{eq:overshoot}. The exponential components were multiplied with a Fermi step function to render the onset.

For every PMT, the sum of these two equations was used to fit the averaged single p.e. waveform to extract the parameters, an example is shown in Fig.~\ref{fig:SPEWaveform}. The fitting was applied to all 192 channels and some important parameters (the averaged values of all 192 PMTs) are listed in Table~\ref{table:parameters}. Some of the parameters were found to follow a Gaussian distribution with a resolution also listed in Table~\ref{table:parameters}. The resolution was also used in the simulation.

\begin{equation}
 \label{eq:log-norm}
 U_{peak}(t) = U^0_{peak}\times exp(-\frac{1}{2}(\frac{ln((t-t_0)/\tau)}{\sigma})^2)
\end{equation}

\begin{equation}
 \label{eq:overshoot}
 \begin{split}
 U_{os}(t) &= U_{peak}(t)\times [U^{Fast}_{os}/(e^{\frac{t_{0}-t}{10ns}}+1)\times exp(-t/\tau_{fast}) \\
  & + U^{Slow}_{os}/(e^{\frac{t_{0}-t}{10ns}}+1)\times exp(-t/\tau_{slow}) \\
  & + U^{Gaus}_{os} \times exp(-\frac{1}{2}(\frac{t-t_0}{\sigma_{os}})^2)]
 \end{split}
\end{equation}

\begin{table}[tb]
  \caption{Parameters for the single p.e. spectrum, which are the averaged values of all 192 PMTs, to be used in the simulation. The amplitude was relative to the main peak if no unit included.}
  \label{table:parameters}
  \begin{tabular}{cc}
    \hline
    \textbf{Parameters} & \textbf{Values}  \\
    \hline
    Main peak amplitude ($U^0_{peak}$) & 42 mV \\
    Peak width ($\tau$) & 8.4$\pm$ 0.3 ns (Gaussian) \\
    Peak shape ($\sigma$) & 0.28 $\pm$ 0.02 (Gaussian) \\
    Fast overshoot amp. ($U^{Fast}_{os}$) & 0.11$\pm$0.02 (Gaussian)\\
    Slow overshoot amp. ($U^{Slow}_{os}$) & 0.03$\pm$0.005 (Gaussian)\\
    Fast overshoot $\tau_{fast}$ & 45 ns\\
    Slow overshoot $\tau_{slow}$ & 290 ns\\
    1$^{st}$ reflection peak amp. & 0.24$\pm$0.02 (Gaussian)\\
    2$^{nd}$ reflection peak amp. & 0.18$\pm$0.02 (Gaussian)\\
    3$^{rd}$ reflection peak amp. & 0.14$\pm$0.015 (Gaussian)\\
    \hline
  \end{tabular}
\end{table}

\section{PMT waveform reconstruction}
\label{Reconstruction}

In Daya Bay, the electronics non-linearity was due to a complex interplay of the LS timing profile, PMT overshoot, and the electronics response. The goal of the newly installed FADC system was to directly measure the current electronics non-linearity, thus the basis of the measurement was to precisely reconstruct PMT charge from the raw waveform. Besides, the system would help us to gain experience of the waveform reconstruction in future LS experiments, such as JUNO\cite{JUNO}.

In this section, we show several frequently used charge reconstruction methods which were examined using MC. The performance was not satisfactory as about a 10\% residual non-linearity was found, which was defined as the ratio of reconstructed charge over the MC true. Then a method based on the deconvolution technique was developed, which was fast, robust, and with a 1\% residual non-linearity.

\subsection{Review of some charge reconstruction algorithms}

There are several commonly used waveform reconstruction algorithms, such as simple integral, CR-(RC)$^n$ shaping and waveform fitting. A detailed investigation was made using these algorithms, and they were found to be unable to deal with the complicated waveform.

\subsubsection{Simple charge integral}

\begin{figure}[!htb]
\begin{centering}
\includegraphics[width=.4\textwidth]{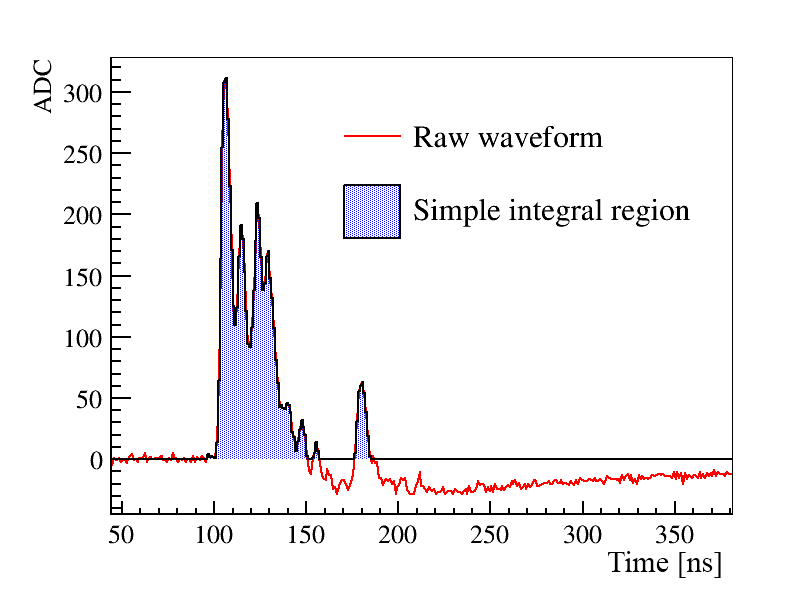}
\caption{\label{fig:SimpleIntegralEx} Example integral region of the simple integral method. The charge of late hits overlying on the overshoot was under-estimated.}
\end{centering}
\end{figure}

Figure~\ref{fig:SimpleIntegralEx} shows an example of the simple charge integral algorithm, which was realized in the following three steps:

1) Determine the hit time crossing the threshold which was about 0.25 p.e.

2) For each hit, integrate the forward and backward region until the waveform returns to the baseline.

3) Divide the integral result by the one of single p.e. to extract the p.e. number.

\begin{figure}[!htb]
\begin{centering}
\includegraphics[width=.4\textwidth]{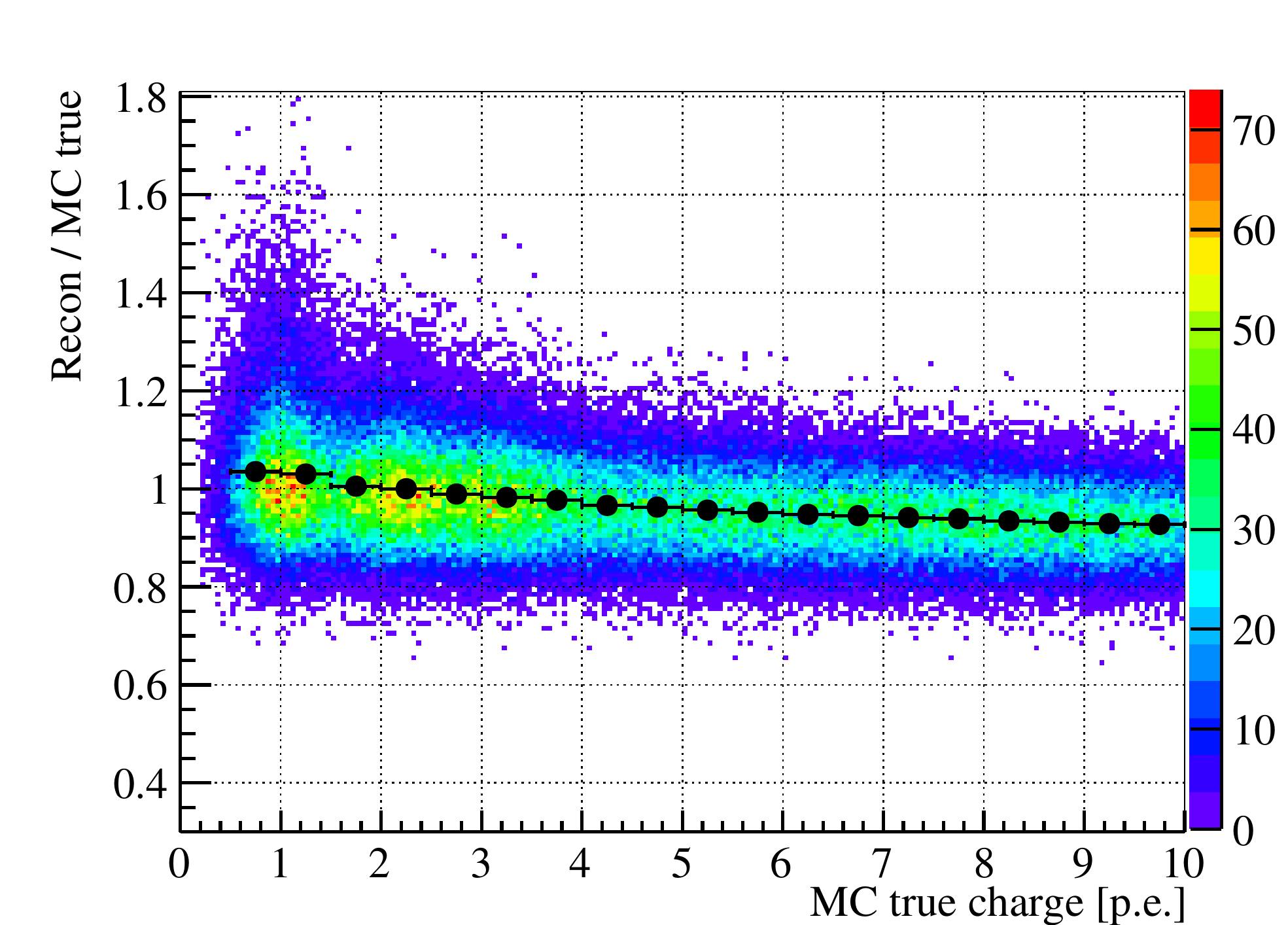}
\caption{\label{fig:SimpleIntegralRe} Residual non-linearity with the simple integral method in MC, which was about 10\% due to the interplay of overshoot and pile-up hits. The color means event number in the certain bin. The X-axis is MC true charge, and the Y-axis is the ratio between the reconstructed and the MC one. The black points show the averaged ratio.}
\end{centering}
\end{figure}

Obviously, due to the late hits overlaying on the overshoot, the simple integral method under-estimated their charge. The electronics simulation described in Sec.~\ref{Simulation} was utilized to examine the residual non-linearity. The results are shown in Fig.~\ref{fig:SimpleIntegralRe}, in which the X-axis is the MC true charge, and the Y-axis is the ratio of the reconstructed charge over the MC one. With the MC true p.e. increasing, the number of late hits was also increasing, inducing more charge under-estimation causing the decreasing trend of the non-linearity curve.

As the simplest charge integral method, this algorithm could be used if the PMT had no overshoot. In the overshoot case, the algorithm was improved in such a way that the baseline of a late hit was re-estimated with the preceding sampling points. However, the improved algorithm still had 3\% residual non-linearity.

\subsubsection{CR-(RC)$^n$ shaping}

The CR-(RC)$^n$ shaping method for charge reconstruction was also investigated. It improved the Signal-to-Noise ratio, but increased the pulse width, induced by pulse pile-up and affected the charge measurement. If the signals' arrival time was concentrated to less than 20 ns, the shaping could give a linear charge measurement. In the LS detector, which had medium and slow scintillation components, the shaping seemed not proper.

\begin{figure}[!htb]
\begin{centering}
\includegraphics[width=.5\textwidth]{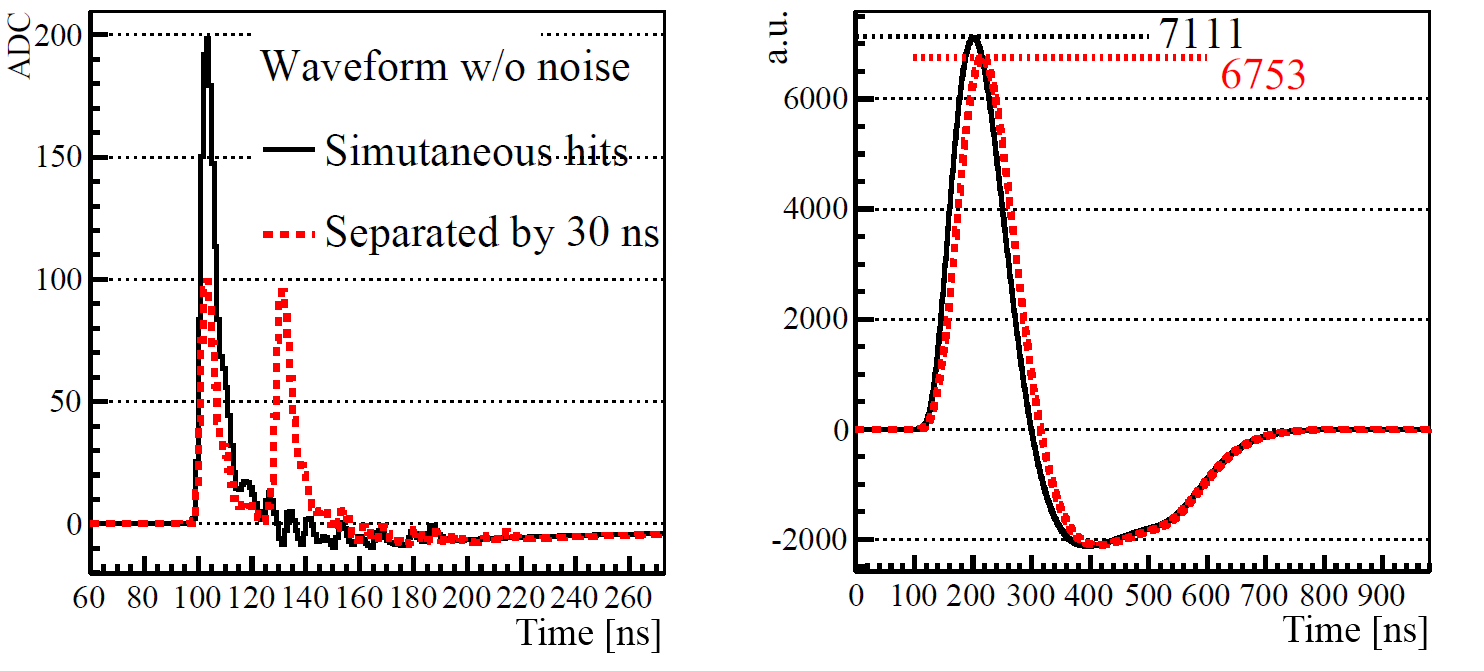}
\caption{\label{fig:CR-RC4-Ex} Left: the black solid line is the simulated waveform of two simultaneous hits and the red dashed line is that of the same two hits separated by 30 ns, both without electronics noise. Right: the two waveforms after CR-(RC)$^4$ shaping. The separated hits have a under-estimated charge, which is 6753 and about 5\% lower than that of simultaneous hits.}
\end{centering}
\end{figure}

\begin{figure}[!htb]
\begin{centering}
\includegraphics[width=.4\textwidth]{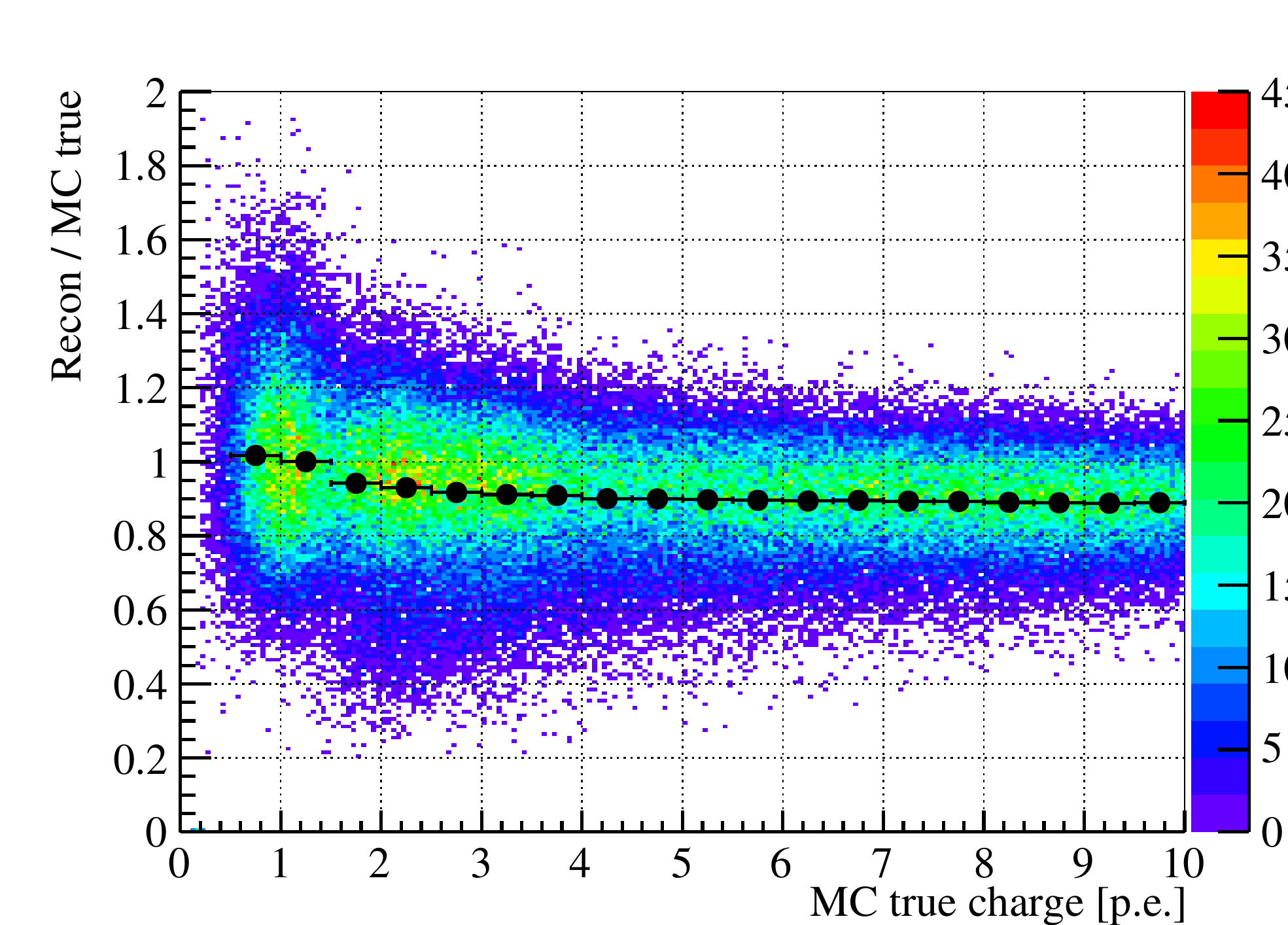}
\caption{\label{fig:CR-RC4-Re} Residual non-linearity of CR-(RC)$^4$ shaping, about 10\%, due to the complex interplay between the shaping constant, overshoot and LS scintillation timing.}
\end{centering}
\end{figure}

Since the current Daya Bay electronics was designed with the CR-(RC)$^4$ shaping circuit, with the $\tau$ of each CR or RC of 25 ns, we chose n=4 as an example . PMT charge was reconstructed with a peak finding algorithm after the CR-(RC)$^4$ circuit, and the integral value was approximated as peak height. As shown in Fig.~\ref{fig:CR-RC4-Ex}, if two hits were separated by 30 ns, the reconstructed charge was under-estimated by 5\% compared to the case of the same two simultaneous hits. With the time separation increasing, the under-estimation was also increasing. With electronics simulation the residual non-linearity was studied and is shown in Fig.~\ref{fig:CR-RC4-Re}, which was also about 10\%.

\subsubsection{Waveform fitting}

\begin{figure}[!htb]
\begin{centering}
\includegraphics[width=.4\textwidth]{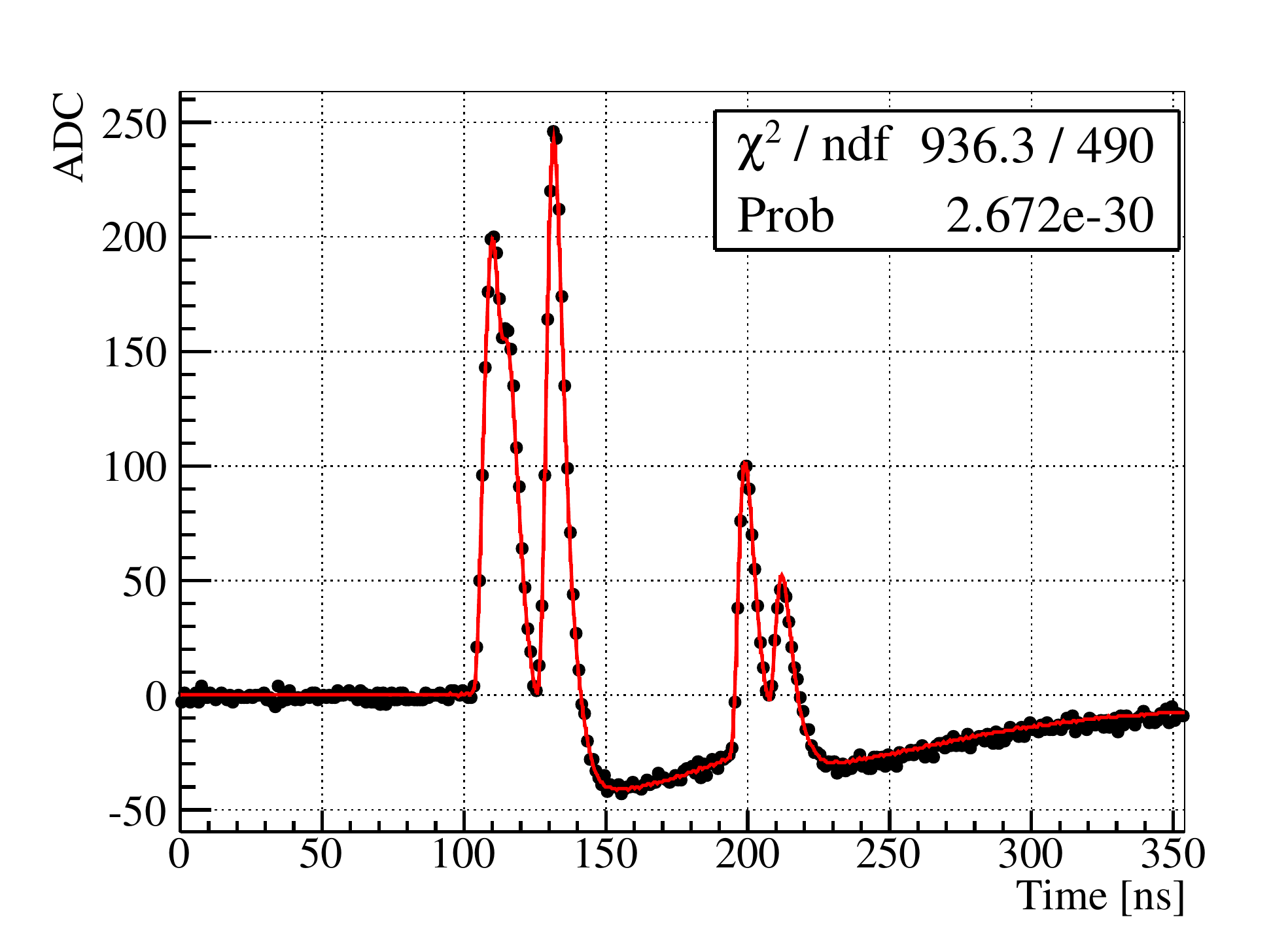}
\caption{\label{fig:Waveform-Fit} An example of waveform fitting. Late hits could be well reconstructed, but the fitting speed was a significant problem with 0.5 s per waveform.}
\end{centering}
\end{figure}

To reconstruct the late hits better, a waveform fitting algorithm was developed, which utilized the calibrated single p.e.~waveform as a template. A fitting example is shown in Fig.~\ref{fig:Waveform-Fit}. Although the residual non-linearity was improved to about 2\%, the algorithm had two shortages to overcome.

The first one was speed. The fitting of one waveform required about 0.5 second, which was a huge workload during the data reconstruction.

The second one was fitting quality. The fitting had a increasing failure rate with increasing number of hits, introducing a residual non-linearity which was difficult to calibrate.

The waveform fitting method could be used in the crosscheck analysis of small event samples, for example, Inverse Beta Decays, etc., in which special care could be taken to examine the fitting quality.

\subsection{A method with the deconvolution technique}

Deconvolution is a well-developed and widely used technique in Digital Signal Processing (DSP)\ \cite{DSP}. It is also utilized in various physics experiments, for example, to extract correct information from pile-up\ \cite{NaI}, pattern recognition\ \cite{Pattern}, and energy spectrum study\ \cite{Sr90}. Furthermore, we found that deconvolution was a powerful tool to reconstruct the PMT charge with good linearity, especially in the case when signals were bi-polar and overlapping.

In the time domain the deconvolution was not easy to process, but in the frequency domain it was rather simple. The raw waveform was converted to the frequency domain with Discrete Fourier Transform (DFT), then multiplied with a noise filter, divided by the frequency response of a calibrated single p.e. waveform, and finally converted to the time domain with Inverse DFT. The DFT and Inverse DFT were done with the Fast Fourier Transform package in the data analysis framework ROOT\ \cite{Root}.

\begin{figure}[!htb]
\begin{centering}
\includegraphics[width=.4\textwidth]{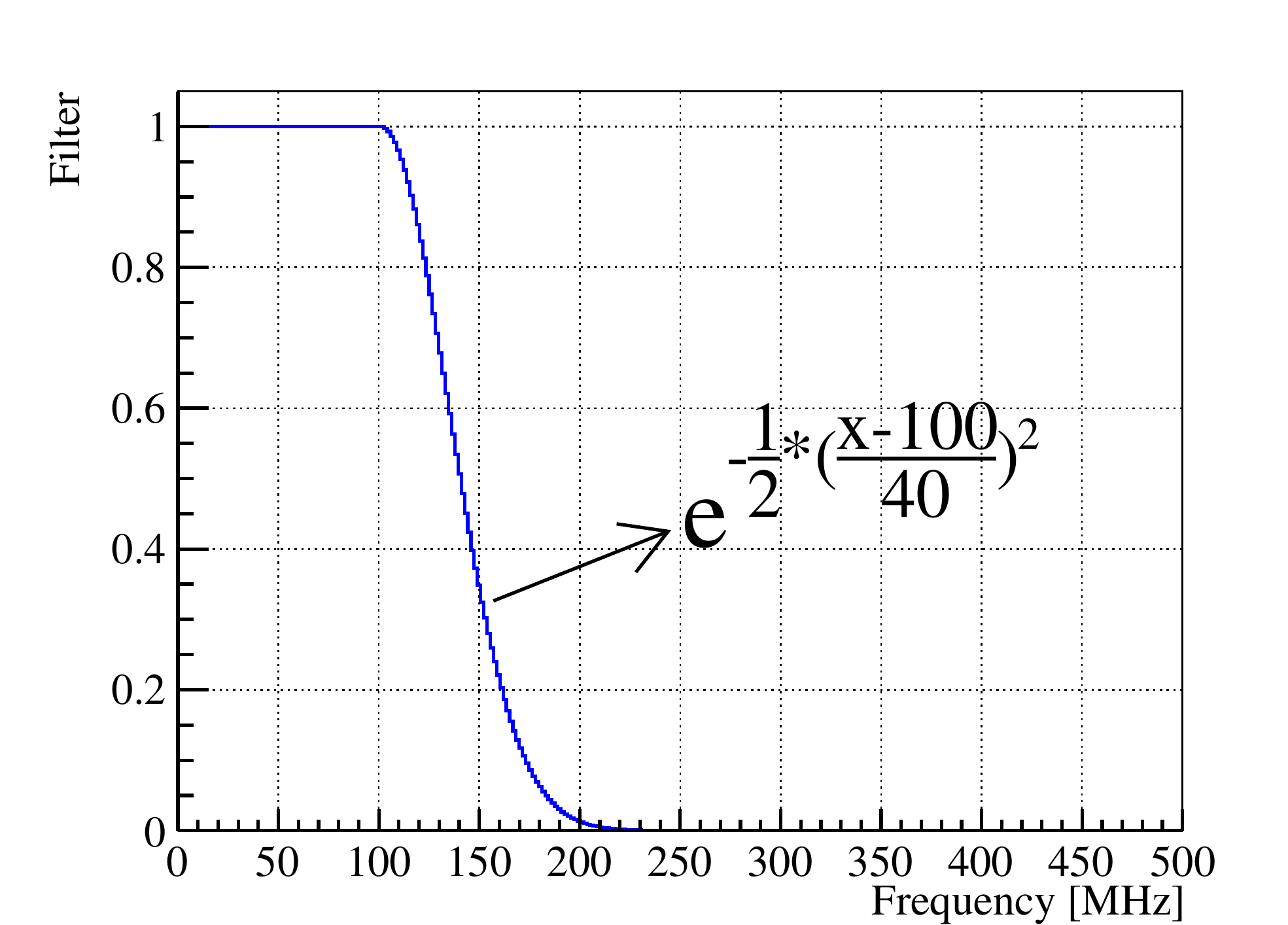}
\caption{\label{fig:FilterEx} The custom-defined low-band pass filter. In case of less than 100 MHz, the filter response equals to 1, while larger than 300 MHz is 0. Between 100 MHz to 300MHz, the filter is described with the Gaussian formula as shown in the plot.}
\end{centering}
\end{figure}

There were a lot of noise filters in the DSP, such as Optimized Wiener Filter, Windowed-Sinc Filter, Gaussian Filter etc\ \cite{DSP}. They were investigated, and a custom-defined low-band pass filter was adopted, as shown in Fig.~\ref{fig:FilterEx}.

\begin{figure}[!htb]
\begin{centering}
\includegraphics[width=.35\textwidth]{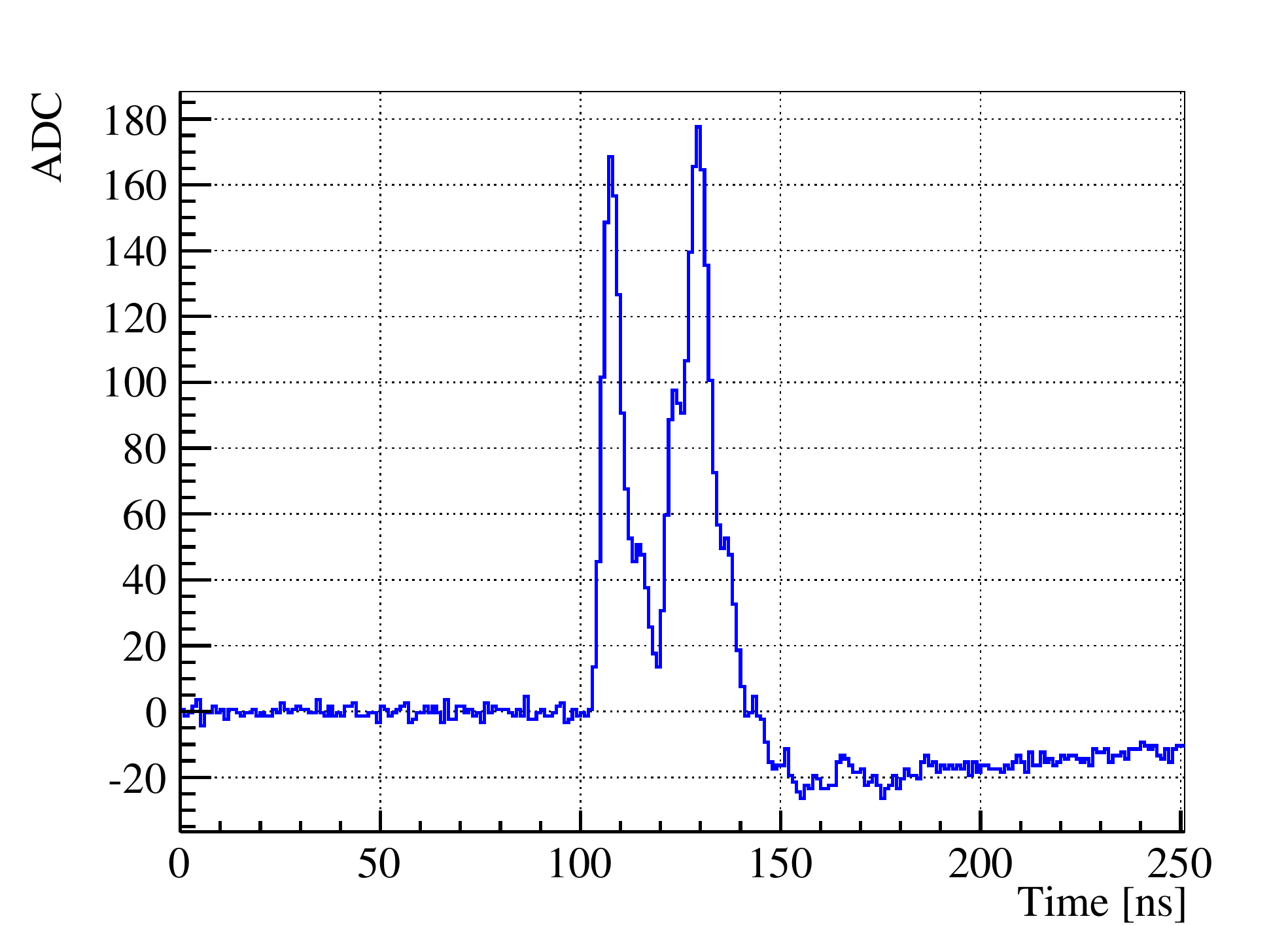}
\includegraphics[width=.35\textwidth]{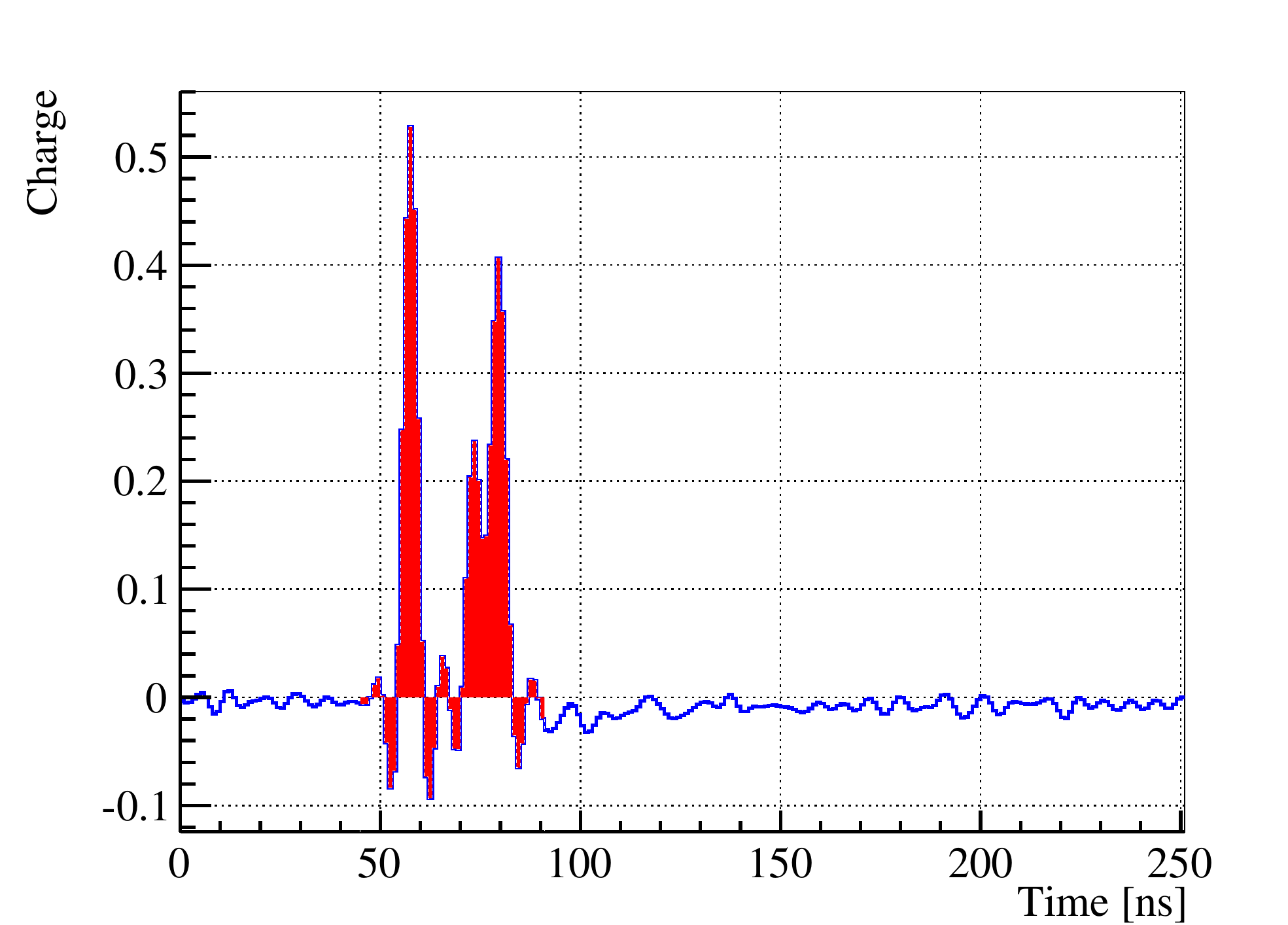}
\caption{\label{fig:DeconvolutionEx} An example of a raw waveform (top) and its deconvolution result (bottom). It can be found that the overshoot has been well removed, but local ringing is introduced by the filter, known as Gibbs effects. To reconstruct the charge linearly, the red region is used to do charge integral, which covers the peak and it's preceding and subsequent ringing for 9 ns.}
\end{centering}
\end{figure}

An example of a raw waveform and its deconvolution is shown in Fig.~\ref{fig:DeconvolutionEx}. During the deconvolution, the peak in spe waveform is forced at 50ns, this is the reason why the deconvoluted peak occurs earlier (50 ns) than the raw waveform. In Fig.~\ref{fig:DeconvolutionEx}, the overshoot was naturally handled, overlapping hits were better separated, and the peak's integral was the p.e. number. However, some local ringing appeared around the peak due to the noise filter's high frequency cut, known as the Gibbs effects\ \cite{Gibbs}. How to extract the PMT charge needs detailed investigation.

It should be noted that the negative ringing could be avoided with an iterative deconvolution method developed by R. Gold\ \cite{Gold}. However, the iteration consumed so much computing time (in general it costs 100 times more computing time) that the Gold deconvolution was not adopted.

\subsubsection{Charge reconstruction}

If the PMT hits arrived at the same time, the local ringing had no influence on the charge reconstruction. However, due to the LS timing profile, the ringing of different hits were overlapped, and a proper integral region should be selected. The red region in Fig.~\ref{fig:DeconvolutionEx} shows an integral example, that each hit integrates the peak and its preceding and subsequent ringing for {\bf x} ns. If two hits were separated by less than {\bf 2*x} ns, the whole region between them was integrated.

\begin{figure}[!htb]
\begin{centering}
\includegraphics[width=.4\textwidth]{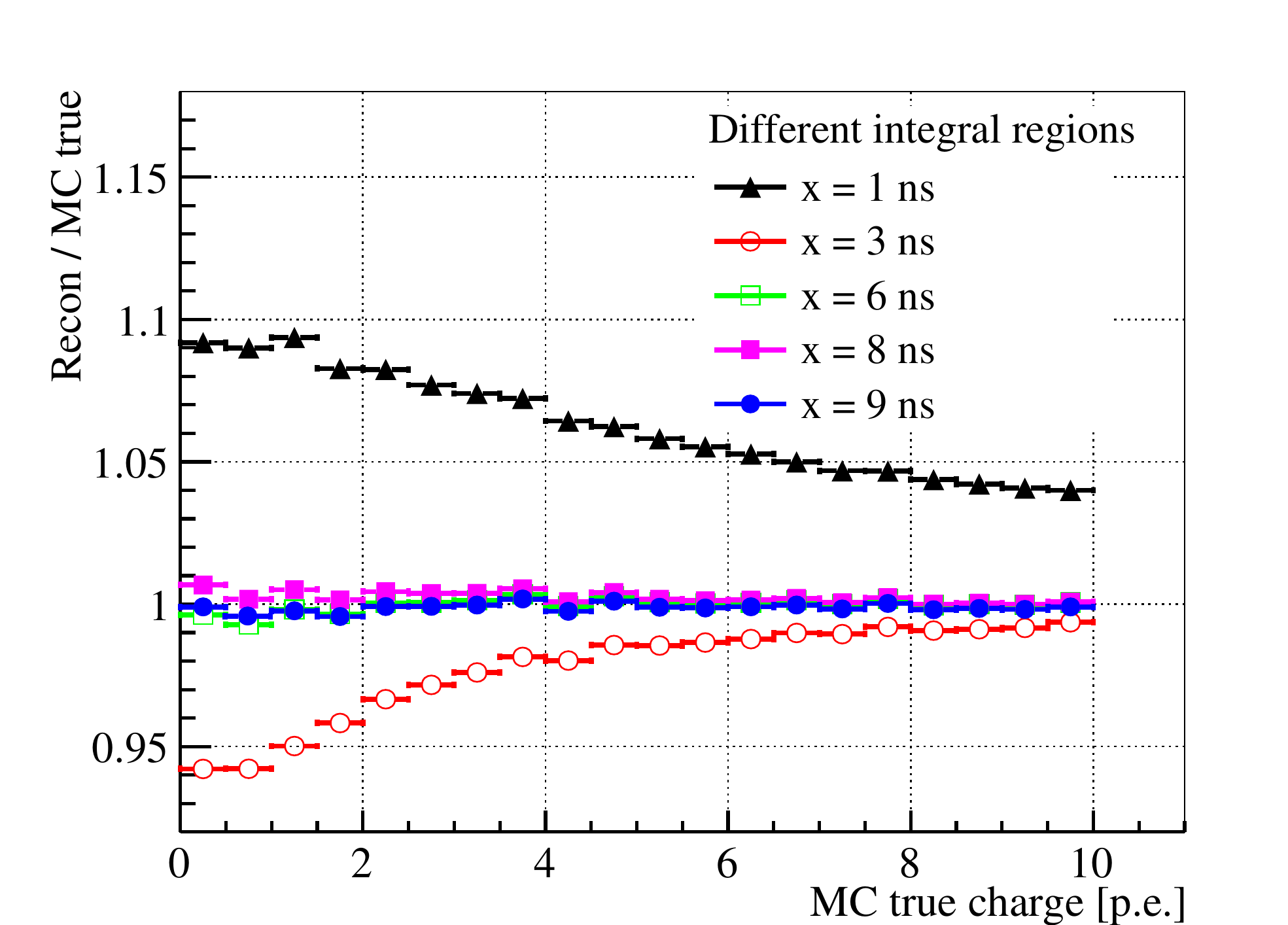}
\caption{\label{fig:DeconvolutionRe} The residual non-linearity of the deconvolution method with different {\bf x}. The blue open circle is {\bf x = 9ns} which has the best performance. }
\end{centering}
\end{figure}

The value of {\bf x} was determined with the electronics simulation. Fig.~\ref{fig:DeconvolutionRe} shows the residual non-linearity with different {\bf x}, and {\bf x = 6, 8, 9} have similar performance. But if we look at their deconvolution results carefully, when {\bf x = 9}, the ringing is well recovered and has minimum effects to charge estimation. In the case {\bf x} equals to 9, the residual non-linearity was smallest and about 1\%, which fulfilled the requirement. Also, charge reconstruction based on the deconvolution technique consumes about 0.5 ms per channel, which is much faster than waveform fitting. The integral region is affected by the ringing, which is introduced by the high frequency cutoff of the noise filter. Using a different filter, one should study the integral region instead of using 9 ns directly.

The increasing and decreasing non-linearity trends with different {\bf x} could be understood. For example, the {\bf x = 0} is expected to only integrate the peak. If there was one hit, only the peak region was used. If there were several hits overlapping, the peak region of one hit also overlapped with the negative ringing of the other hits. Then the ratio between reconstructed charge and true charge was smaller than in the one hit case, introducing the decreasing trend. In the bottom plot of Fig.~\ref{fig:DeconvolutionEx}, we can see the ringing is go to finish since 6 ns, thus integrate 6, 7, 8 and 9 ns give similar results, as shown in Fig.~\ref{fig:DeconvolutionRe}.

Since the value of {\bf x} was determined with MC, to validate the results, MC are compared with data, as described in the next section.

\subsection{Summary of the reconstruction algorithms}

The interplay between the LS timing profile and PMT overshoot would degrade the performance of two charge reconstruction methods: simple integral and CR-(RC)$^4$ shaping. To pick up the late hits overlaying on the overshoot of earlier hits, the waveform fitting method was tested, and found too slow to be used in the reconstruction for large data samples. Then an algorithm with the deconvolution technique was developed. The performance of these reconstruction algorithms is summarized in Table ~\ref{table:comparison}. It was found that the deconvolution method is the most powerful tool to deal with the overshoot and overlapping, and has the smallest residual non-linearity.

Besides the charge measurement, the algorithms had different timing separation abilities for pile-up hits. The waveform fitting and deconvolution could discriminate hits separated larger than 10 ns, while for the simple integral method it was 20 ns and for the Daya Bay CR-(RC)$^4$ shaping 40 ns.

\begin{table*}[!htb]
  \caption{The summary table for different charge reconstruction algorithms.}
  \label{table:comparison}
  \begin{tabular}{ccccc}
    \hline
    Algorithms & Speed per channel & Robustness & Residual non-linearity & Pile-up hits separation\\
    \hline
    Simple integral & less than 0.1 ms & No failure& 3\% to 10\% & Larger than 20 ns\\
    \hline
    CR-(RC)$^4$ & 0.2 ms & No failure& 10\% & Larger than 40 ns\\
    \hline
    Waveform fitting & 0.5 s & \tabincell{c}{Sometimes fails and \\ difficult to define failure} & 2\% & Larger than 10 ns\\
    \hline
    Deconvolution & 0.5 ms & No failure & 1\% & Larger than 10 ns \\
    \hline
  \end{tabular}
\end{table*}

\section{Comparison between electronics simulation and data}
\label{Validation}

Since the algorithm performance was studied with MC, to validate the conclusions, MC should be compared with data. The most direct way was to compare the residual non-linearity between data and MC, but in the data there was no true p.e.~information.  As such, a ratio was defined where the deconvolution result with {\bf x=9 ns} was used as the denominator, and the results from other methods were used as the numerator. The ratio was compared between data and MC, and 1\% agreements were achieved, indicating confidence in the MC simulation.

\subsection{Comparison of different charge integral methods for the deconvolution algorithm}

After deconvolution, how to integrate the local ringing was studied with MC, and a best solution was given which covered the preceding and subsequent 9 ns. The MC studies also showed that integrating different regions had different non-linearity, which was expected to be repeated in data.

\begin{figure}[!htb]
\begin{centering}
\includegraphics[width=.4\textwidth]{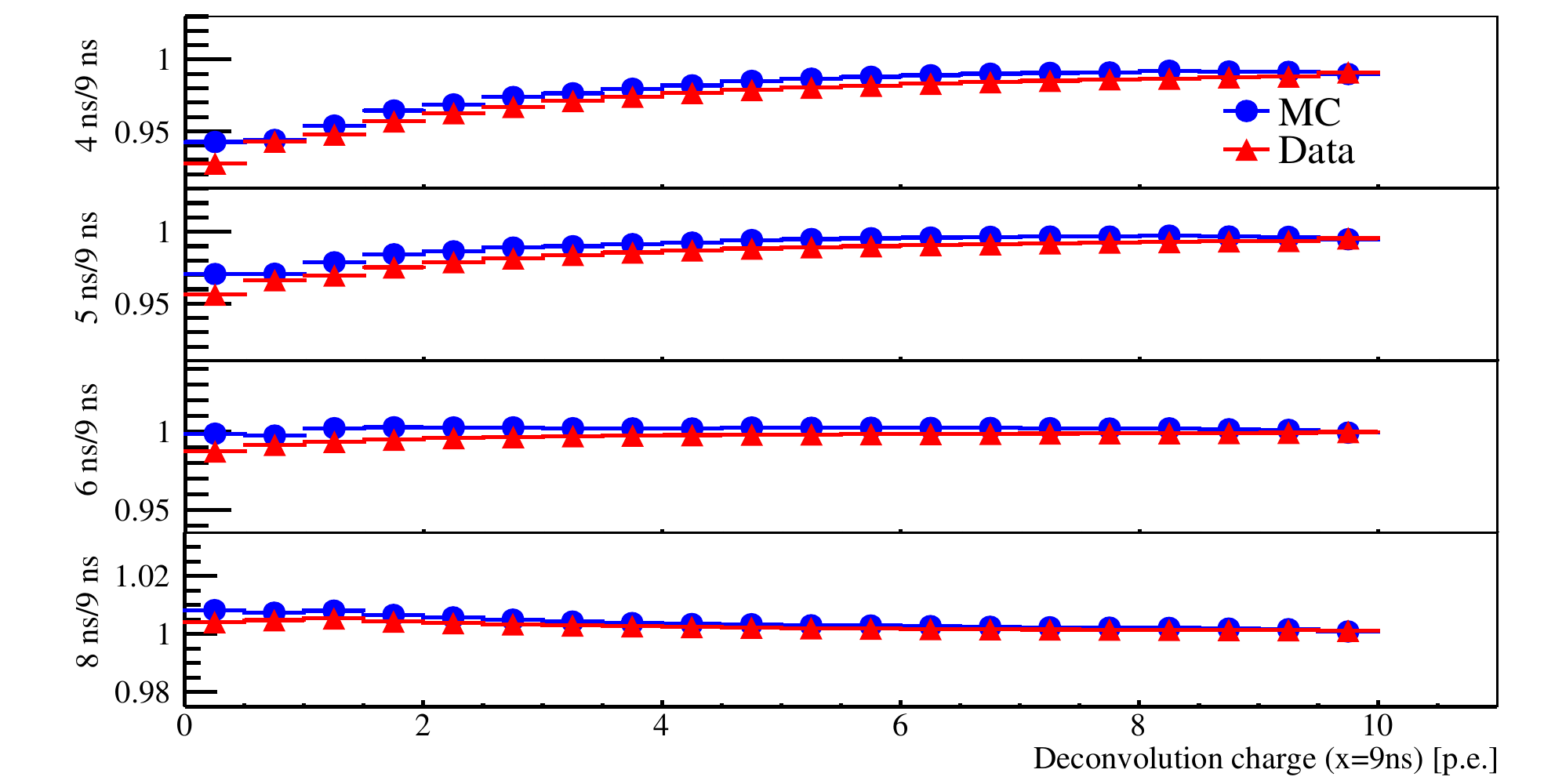}
\caption{\label{fig:DeconvolutionComparison} Ratio of the reconstructed charge between different charge integral regions after deconvolution. Data and MC agree to 1\%, indicating the MC has well described the waveform and the deconvolution method. The four pads from top to bottom show the ratio of {\bf (x = 4, 5, 6, 8 ns)/(x = 9 ns)} respectively. }
\end{centering}
\end{figure}

A ratio was defined as $\frac{Deconvolution~(x~=~a~ns)}{Deconvolution~(x~=~9~ns)}$, where {\bf x} meant the integral region, and {\bf a} ran from 0 to 8. Some comparison results are shown in Fig.~\ref{fig:DeconvolutionComparison} as examples. Data and MC agree to 1\%, indicating the MC has well described the waveform and the deconvolution method.

\subsection{Comparison of different FADC readout window lengths}

The Discrete Fourier Transform assumes that the finite sequence of sample points is periodical. In the waveform reconstruction, if the overshoot was not well recovered at the end of the readout window, the periodical assumption would treat the un-recovered baseline as a jump. Thus an additional non-linearity would be induced, and the larger the jump was, the larger non-linearity we got.

\begin{figure}[!htb]
\begin{centering}
\includegraphics[width=.4\textwidth]{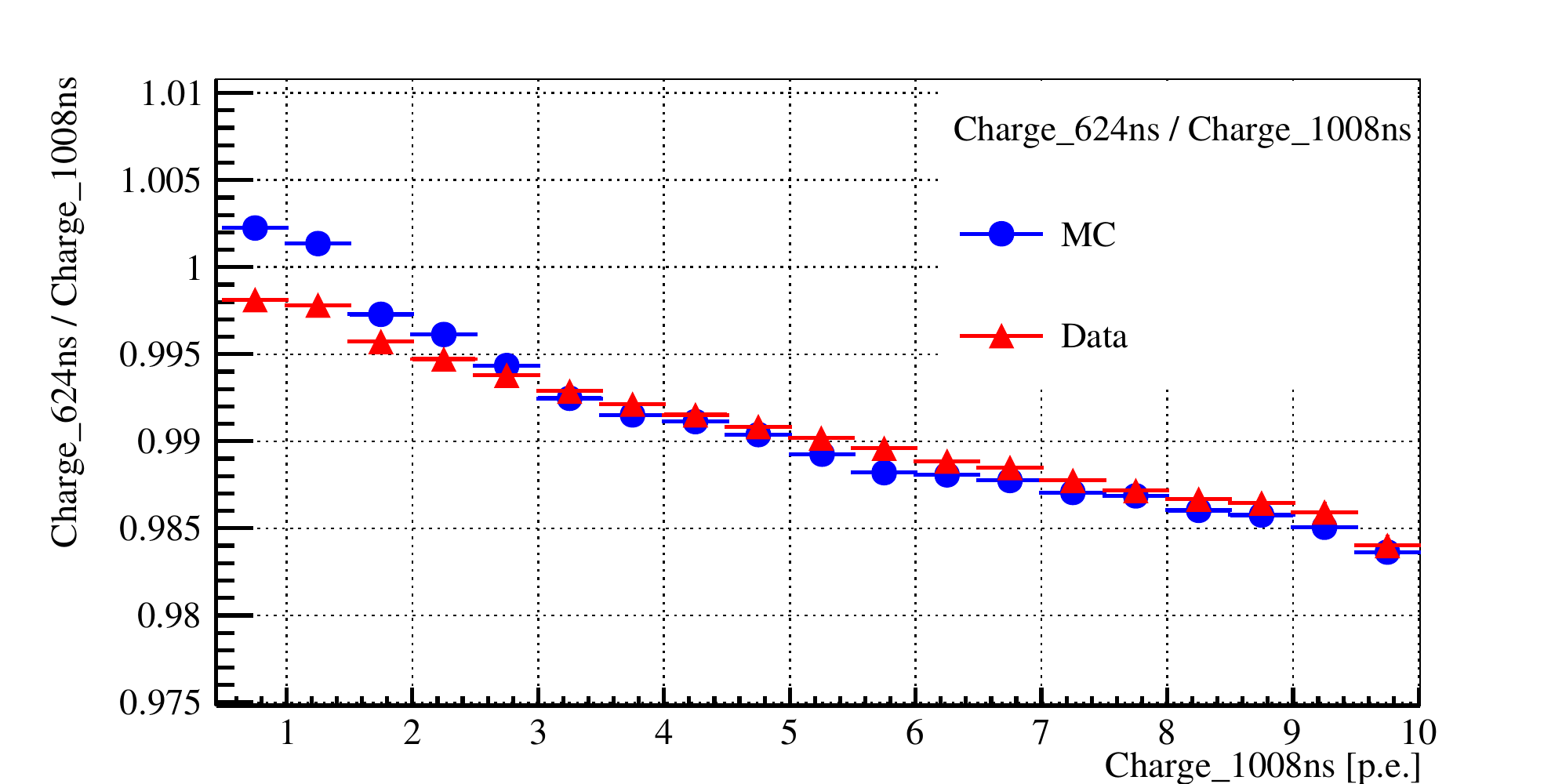}
\caption{\label{fig:compare624vs1008} Ratio of the reconstructed charge between different FADC readout window lengths. MC and data agree to less than 1\%, and the differences between 624 ns and 1008 ns are due to the coupling of DFT's periodical assumption and the long overshoot.}
\end{centering}
\end{figure}

The Daya Bay FADC readout window length was set to 1008 ns and the previous results were performed with this length. To validate the additional non-linearity, the first 624 ns was used to reconstruct the PMT charge and compared to the one reconstructed from the full waveform. For consistency, both used chose hits with hit time less than 500 ns for charge reconstruction. As shown in Fig.~\ref{fig:compare624vs1008}, both data and MC showed a decreasing trend and they agreed to better than 1\%.

\subsection{Comparison of different reconstruction algorithms}

The simple integral and CR-(RC)$^4$ methods were more sensitive to the PMT waveform features, such as overshoot, reflections and hit time profiles. Comparison between them and the deconvolution method could validate the description of waveform features in MC.

\begin{figure}[!htb]
\begin{centering}
\includegraphics[width=.4\textwidth]{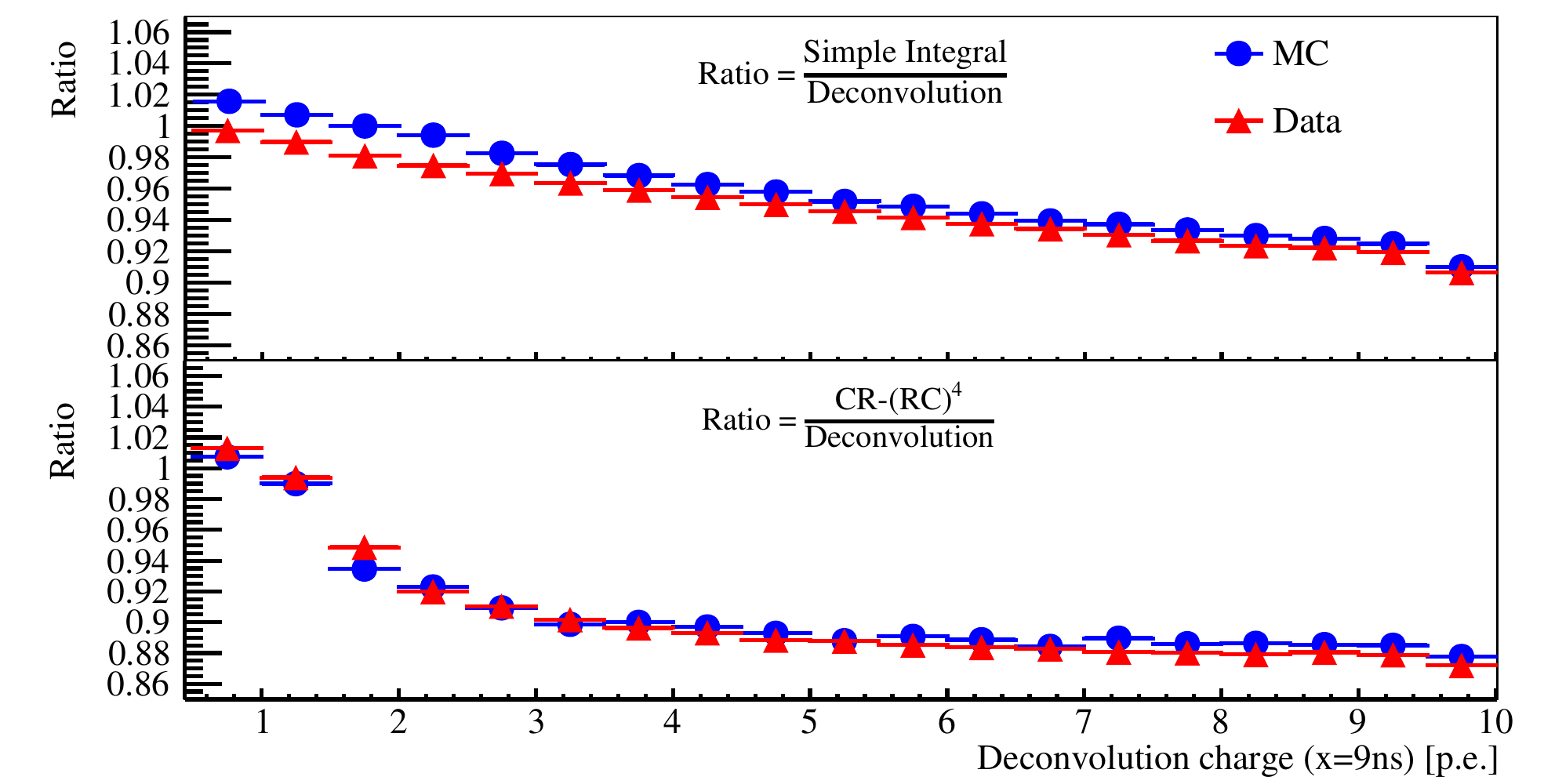}
\caption{\label{fig:SICRvsDe} Top: comparison between the simple integral method and the deconvolution one; bottom: between the CR-(RC)$^4$ shaping and the deconvolution one. Most of the points agree to 1\% indicating the MC has described the waveform features well.}
\end{centering}
\end{figure}

Two ratios were defined as $\frac{Simple charge integral}{Deconvolution~(x~=~9~ns)}$ and $\frac{CR-(RC)^4}{Deconvolution~(x~=~9~ns)}$. Data and MC comparison results are shown in Fig.~\ref{fig:SICRvsDe} and most of them agree to 1\%, indicating the MC has described the waveform features well.

\section{Summary}
\label{Summary}

To directly measure the current electronics (FEE) non-linearity, the Daya Bay experiment installed a Flash ADC system on AD1. The system has been stably running since Feb. 2016, and no influence was found to the FEE. The scientific goal required a precise PMT charge reconstruction algorithm, and several waveform reconstruction methods were developed and examined. The one based on the deconvolution technique was found to have the best performance, with a 1\% residual non-linearity. A confident electronics simulation was developed and it agreed with data to 1\%. In particular, the FADC data, combined with the developed deconvolution charge reconstruction method validate the electronics non-linearity modeling currently implemented in Daya Bays oscillation~\cite{DYB-Oscillation-1230days} ~\cite{DYB-Oscillation} and reactor physics analyses~\cite{DYB-Reactor}.

This work has been supported by the National Natural Science Foundation of China (11390385).




\end{document}